\documentclass[iop, apjl, twocolappendix, twocolumn]{aastex631}
\usepackage{graphicx}
\usepackage{bm}
\usepackage{amsmath,amssymb}
\usepackage{natbib}
\usepackage{booktabs}
\usepackage{color}
\usepackage{units}

\usepackage[normalem]{ulem}
\newcommand{\tfb}{\ensuremath{t_{\rm fb}}}
\newcommand{\program}[1]{\textsc{#1}}

\defcitealias{Metzger22}{M22}

\newcommand{\be}{\begin{equation}}
\newcommand{\ee}{\end{equation}}

\begin{document}
\title{Tidal Disruption Events through the Lens of the Cooling Envelope Model}

\author{Nikhil Sarin}
\affil{Nordita, Stockholm University and KTH Royal Institute of Technology \\
Hannes Alfvéns väg 12, SE-106 91 Stockholm, Sweden}
\affil{Oskar Klein Centre for Cosmoparticle Physics, Department of Physics,
Stockholm University, AlbaNova, Stockholm SE-106 91, Sweden}

\author{Brian D.~Metzger}
\affil{Department of Physics and Columbia Astrophysics Laboratory, Columbia University, New York, NY 10027, USA}
\affil{Center for Computational Astrophysics, Flatiron Institute, 162 5th Ave, New York, NY 10010, USA} 

\begin{abstract}

The cooling envelope model for tidal disruption events (TDE) postulates that while the stellar debris streams rapidly dissipate their bulk kinetic energy (``circularize"), this does not necessarily imply rapid feeding of the supermassive black hole (SMBH). The bound material instead forms a large pressure-supported envelope which powers optical/UV emission as it undergoes gradual Kelvin-Helmholtz contraction. We present results interpreting a sample of 15 optical TDE within the cooling envelope model in order to constrain the SMBH mass $M_{\rm BH}$, stellar mass $M_{\star}$, and orbital penetration factor $\beta$. The distributions of inferred properties from our sample broadly follow the theoretical expectations of loss-cone analysis assuming a standard stellar initial mass function. However, we find a deficit of events with $M_{\rm BH} \lesssim 5\times 10^{5}M_{\odot}$ and $M_{\star} \lesssim 0.5M_{\odot}$, which could result in part from the reduced detectability of TDEs with these properties. Our model fits also illustrate the predicted long delay between the optical light curve peak and when the SMBH accretion rate reaches its maximum. The latter occurs only once the envelope contracts to the circularization radius on a timescale of months to years, consistent with delayed-rising X-ray and non-thermal radio flares seen in a growing number of TDE.


\end{abstract}

\section{\label{sec:intro}Introduction}
Wide-field time-domain surveys such as the Zwicky Transient Facility (ZTF; \citealt{Bellm+19}) are rapidly growing the observed populations of several distinct classes of electromagnetic transients. One such class are tidal disruption events (TDE), which occur when a star in a galactic nucleus approaches the central supermassive black hole (SMBH) sufficiently close to be torn apart by tidal forces \citep{Hills75,Luminet&Carter86,Rees88,Evans&Kochanek89,Stone+13,Guillochon&RamirezRuiz13,Coughlin&Nixon22}.  These luminous events offer unique insights into a wide range of topics in high-energy astrophysics, including the dynamics (e.g., \citealt{Magorrian&Tremaine99,Ivanov+05}) and demographics (e.g., \citealt{Kochanek16a,Gallegos-Garcia+18,Mockler+22}) of stellar populations in galactic nuclei, the otherwise challenging to probe low-end tail of the SMBH mass distribution (e.g., \citealt{Stone&Metzger16,Wevers+17,Ryu+20,Zhou+21}), strong-field features of general relativity (e.g., \citealt{Kesden12,Stone&Loeb12,Guillochon&RamirezRuiz15,Lu+17,Ryu+20b,Mummery&Balbus20}), and the conditions for relativistic jet formation (e.g., \citealt{Giannios&Metzger11,Krolik&Piran12,DeColle+12,Tchekhovskoy+14,Lu+17}).

Upon their discovery in UV (e.g., \citealt{Stern+04,Gezari+06}) and optical (e.g., \citealt{vanVelzen+11,Cenko+12,Arcavi+14}) surveys, several features of observed TDE flares were unexpected, particularly their high optical luminosities $L \gtrsim 10^{43}$ erg s$^{-1}$, modest effective temperatures $T_{\rm eff} \approx 10^{4.2}-10^{4.7}$ K, and correspondingly large photosphere radii $\approx 10^{14}-10^{15}$ cm (e.g., \citealt{Arcavi+14,Holoien+14,Hung+17,vanVelzen+21}; see \citealt{Gezari21} for a recent review). It was generally assumed prior to these discoveries that the disrupted stellar debris would form a compact accretion disk around the SMBH, comparable in size to the tidal sphere (typically tens or hundreds of gravitational radii), thus generating most of its blackbody emission at soft X-ray energies rather than in the optical/UV bands (e.g., \citealt{Rees88,Cannizzo+90,Lodato&Rossi11}).

After reaching peak luminosity, some TDE light curves decay roughly as $L \propto t^{-5/3}$ (e.g., \citealt{Gezari+06,Hung+17}), consistent with the predicted mass fall-back rate for complete disruptions \citep{Phinney89,Guillochon&RamirezRuiz13}. This has motivated semi-phenomenological light curve models like \texttt{MOSFiT} which assume the radiated luminosity to scale with the fall-back rate (e.g., \citealt{Guillochon+18,Mockler+19,Nicholl+20,Coughlin&Nicholl23}).  However, the post-maximum decay is sometimes better fit as an exponential (e.g., \citealt{Holoien+16,Blagorodnova+17}) or may even exhibit a flattening or secondary peaks (e.g.,~\citealt{Leloudas+16,vanVelzen+19,Wevers+19}). Furthermore, while thermal X-ray emission consistent with a compact accretion disk is detected from some optically-selected TDEs, the rise of the X-ray  (e.g., \citealt{Gezari+06,Gezari+17,Kajava+20,Yao+22,Liu+22}) and radio (e.g., \citealt{Alexander+20,Horesh+21,Horesh+21b,Cendes+22}) light curves are often delayed by months or even years after the optical emission has peaked.

\begin{figure*}[ht!]
\centering
  \begin{tabular}{cc}     
        \includegraphics[width=1.0\textwidth]{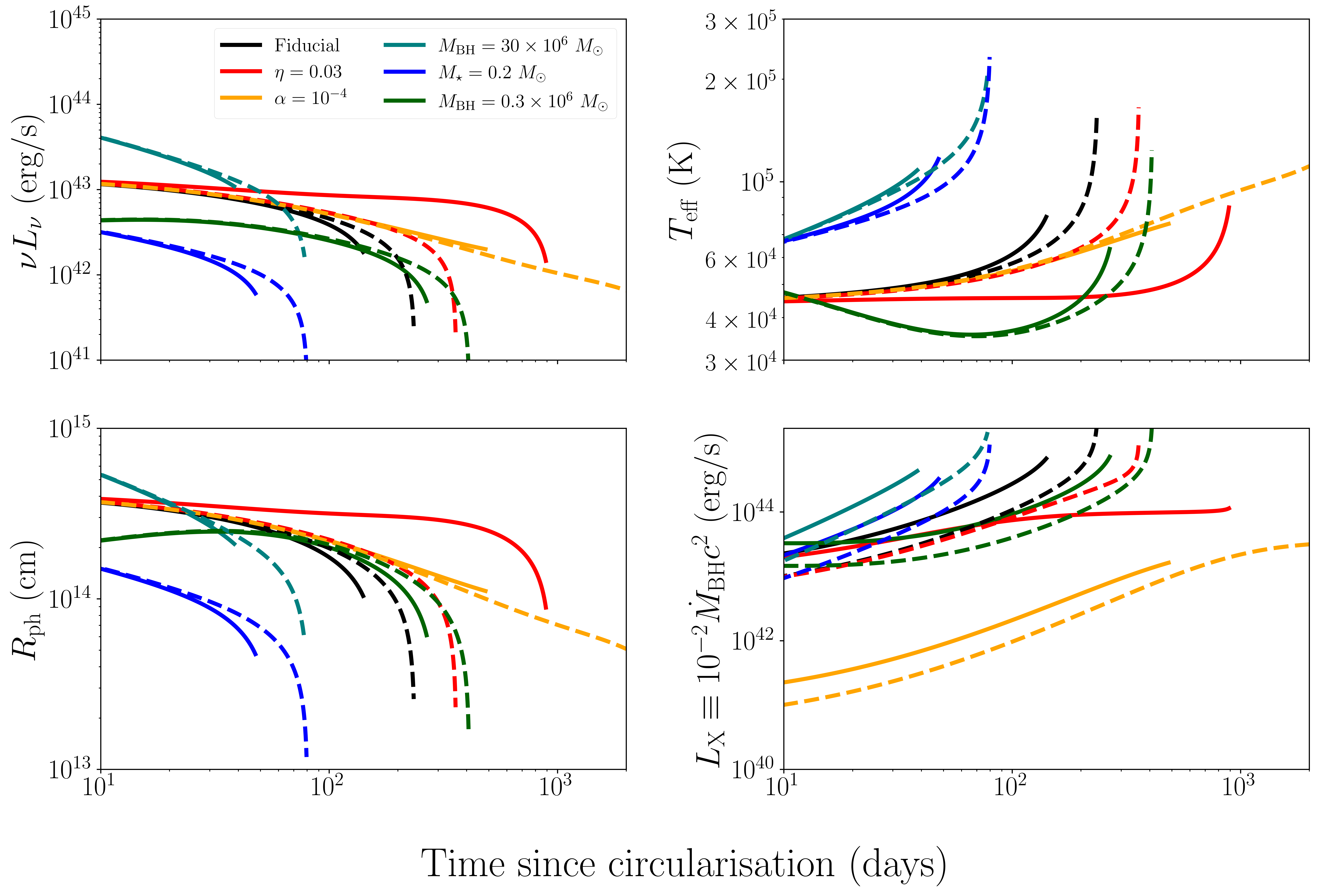} \hspace{-.4cm}
  \end{tabular}
  \caption{Evolution of $\nu L_{\nu}$ light curve at $\nu = \unit[6\times 10^{14}]{Hz}$, the effective temperature, photospheric radius, and proxy X-ray lightcurves $L_{\rm X} \equiv 10^{-2}\dot{M}c^{2}$ (where $\dot{M}$ is the SMBH accretion rate) of the post-circularization cooling envelope emission phase for different parameter choices. The black curves correspond to a fiducial model with SMBH mass $M_{\rm{BH}} = \unit[2\times 10^6]{M_{\odot}}$, stellar mass $M_{\star} = \unit[1]{M_{\odot}}$, penetration factor $\beta =1$, disk viscosity $\alpha=0.01$, and SMBH feedback efficiency parameter $\eta=10^{-4}$.  Solid-colored curves show the effect of changing various parameters as marked while keeping the other parameters fixed at the fiducial model values. The dashed curves correspond to $\beta = 5$ with the rest of the parameters the same as the corresponding solid-coloured curve. The sudden decrease in the photosphere radius and concomitant increase (decrease) in the emission luminosity (temperature) corresponds to the final contraction of the envelope to the circularization radius. While this signals the end of the envelope phase, thermal and non-thermal emission from the accretion disk and/or its jet can fully commence after this transition. 
  } 
\label{fig:fiducial}  
\end{figure*}

These unexpected features have motivated alternative models to power TDE UV/optical light curves rather than standard disk emission. Simulations of the disruption and its aftermath (e.g., \citealt{Hayasaki+13,Guillochon&RamirezRuiz13,Shiokawa+15,Bonnerot+16,Hayasaki+16,Sadowski+16,Steinberg+19,Bonnerot&Lu20,Ryu+21,Andalman+22,Steinberg&Stone22}) which aim to determine the processes by which the debris circularizes have focused on collisions between the bound debris streams (e.g., \citealt{Hayasaki+13,Shiokawa+15,Lu&Bonnerot20}) and how they are hastened or delayed by effects such as radiative cooling or general-relativistic precession (e.g., \citealt{Dai+15,Guillochon&RamirezRuiz15,Andalman+22,Bonnerot&Stone21, Wong2022}). Motivated by the finding that debris circularization can be delayed by many orbits, it was proposed that TDE light curves are powered by stream-stream collisions (e.g., \citealt{Piran+15,Ryu+20}).  On the other hand, some recent simulations find rapid circularization even when precession effects are unimportant \citep{Steinberg&Stone22}. If a modest fraction of the fallback material reaches small scales around the SMBH, the resulting accretion could produce observed UV/optical emission via reprocessing by radially-extended material (e.g., \citealt{Strubbe&Quataert09,Guillochon&RamirezRuiz13,Miller15,Metzger&Stone16,Roth+16,Dai+18,Wevers+19,Lu&Bonnerot20}).

\citet[hereafter \citetalias{Metzger22}]{Metzger22} present a model for TDE emission which assumes circularization is indeed rapid but {\it not} that this necessarily implies equally rapid SMBH feeding. The essential idea is that, due to the weak binding energy of the stellar debris relative to its angular momentum imparted by the disruption process (``virial'' radius $\gg$ ``circularization'' radius), and the inability of super-Eddington fallback to cool radiatively, the virialized debris will form a hot quasi-spherical pressure-supported envelope, rather than most of the material residing in a rotationally-supported disk (see also \citealt{Loeb&Ulmer97,Coughlin&Begelman14}). Although the incorporation of fall-back material heats the envelope, its luminosity and temperature evolution are mainly driven by radiative cooling and gradual envelope contraction, as occurs (at least initially) on the Kelvin-Helmholtz timescale. The inner boundary of the radially-extended envelope attaches smoothly onto a Keplerian disk near the much smaller circularization radius, which then acts as a sink of mass onto the SMBH and a source of accretion energy which can delay the envelope contraction.  

\citetalias{Metzger22} show that the overall timescales and shapes of the predicted light curve decay qualitatively match some TDE observations, such as the predicted power-law decay $\nu L_{\nu} \propto t^{-3/2}$ for a passively cooling envelope which is (coincidentally) similar to the canonical $t^{-5/3}$ fall-back. The model also naturally predicts the observed physical scale and evolution of the photosphere radius, which are essentially free parameters in fall-back powered models such as \texttt{MOSFiT} \citep{Guillochon+18,Nicholl+22}.  Because the rise in the SMBH accretion rate is delayed by the envelope contraction, the model accounts for the long observed lag between the optical peak and other physical processes (e.g., disk UV and X-ray emission, jetted radio emission) which instead more directly traces accretion onto the SMBH from smaller radii near the event horizon.

In this paper, we present results from fitting a sample of UV/optical TDE flares with the \citetalias{Metzger22} cooling envelope model. We briefly describe the model and present our results for a sample of TDE in Sec.~\ref{sec:model} and~\ref{sec:results}, respectively. We discuss implications of our results in Sec.~\ref{sec:discussion}, followed by a brief summary of our conclusions in Sec.~\ref{sec:conclusion}.

\section{Model}
\label{sec:model}
The \citetalias{Metzger22} model evolves the envelope mass $M_{e}$ and virial radius $R_{v}$, accounting for various sources/sinks of mass and energy, respectively. The envelope is assumed to possess a radial density profile $\rho \propto r^{-\xi}$ from the circularization radius $R_{\rm c} = 2R_{\rm t}/\beta$ to $R_{v}$ (and an exponential cutoff at $r > R_{v}$), where $\beta \equiv R_{\rm t}/R_{\rm p}$ is the orbital penetration factor of the disrupted star, and $R_{\rm p}$ and $R_{\rm t}$ are the orbital pericenter radius and tidal radius, respectively. We take $\xi = 1$ throughout this work.

The bolometric luminosity $L_{\rm rad} = L_{\rm edd} + L_{\rm fb}$ of the envelope is comprised of the Eddington luminosity $L_{\rm edd}$ of the SMBH of mass $M_{\rm BH}$ which supports the envelope in hydrostatic equilibrium and that due to deposition by fall-back accretion, $L_{\rm fb} \propto \dot{M}_{\rm fb},$ where $\dot{M}_{\rm fb}(t) \propto t^{-5/3}$ is the rate of mass fall-back. The envelope's effective temperature and photosphere radius obey
\begin{equation}
T_{\rm eff} = \left(\frac{L_{\rm rad}}{4\pi \sigma R_{\rm ph}^{2}}\right)^{1/4}; R_{\rm ph} = R_{v}\left(1 + \ln\left(\frac{\kappa_{\rm es} M_{e}}{10\pi R_{v}^2}\right)\,\right),
\label{eq:Teff}
\end{equation}
where $\kappa_{\rm es} = 0.35$ cm$^{2}$ g$^{-1}$ is the electron-scattering opacity.

Fig.~\ref{fig:fiducial} shows several example optical light curves, which demonstrate typical features of the cooling envelope evolution and highlight some scalings. In particular, we show the evolution of the photospheric radius and effective temperature alongside the $\nu L_{\nu}$ light curve at $\nu = \unit[6\times 10^{14}]{Hz}$ for a few representative parameter choices. The photosphere radius contracts as the envelope radiates energy and becomes more tightly bound to the SMBH, causing the effective temperature to rise (initially gradually) in time. The initial luminosity and decay time of the light curve depend primarily on the Kelvin-Helmholtz timescale of the envelope, 
\be
t_{\rm KH} = \frac{2GM_{\rm BH}M_{e}}{5L_{\rm edd}R_{v,0}} \approx 24\,{\rm d}\,m_{\star}^{13/15}M_{\rm BH,6}^{-2/3}\left(\frac{R_v}{R_{v,0}}\right)^{-1},
\label{eq:tKH}
\ee
where $R_{v,0}$ is the initial envelope radius set by the binding energy imparted during the TDE, $M_{\rm BH} = 10^{6}M_{\rm{BH, 6}}M_{\odot}$ is the SMBH mass, and $M_{\star} = m_{\star}M_{\odot}$ is the mass of the disrupted star mass. We follow \citetalias{Metzger22} in adopting a mass-radius relationship $m_{\star} \propto (R_{\star}/R_{\odot})^{4/5}$ appropriate to lower-main sequence star (e.g., \citealt{Kippenhahn&Weigert90}), which likely constitute the majority of TDE victims. However, as the envelope contracts $R_v \propto t^{-1}$ approaching the circularization radius ($R_{v} \rightarrow R_{\rm c}$), the accretion rate onto the SMBH rises and the shape of the light curve shape becomes sensitive to the SMBH feedback efficiency and disk viscosity. Both effects are also seen in the evolution of the photospheric radius, consistent with the scalings described in \citetalias{Metzger22} and summarised in Appendix \ref{sec:scalings}. 

\begin{figure*}[ht!]
\centering
  \begin{tabular}{cc}             \includegraphics[width=0.87\textwidth]{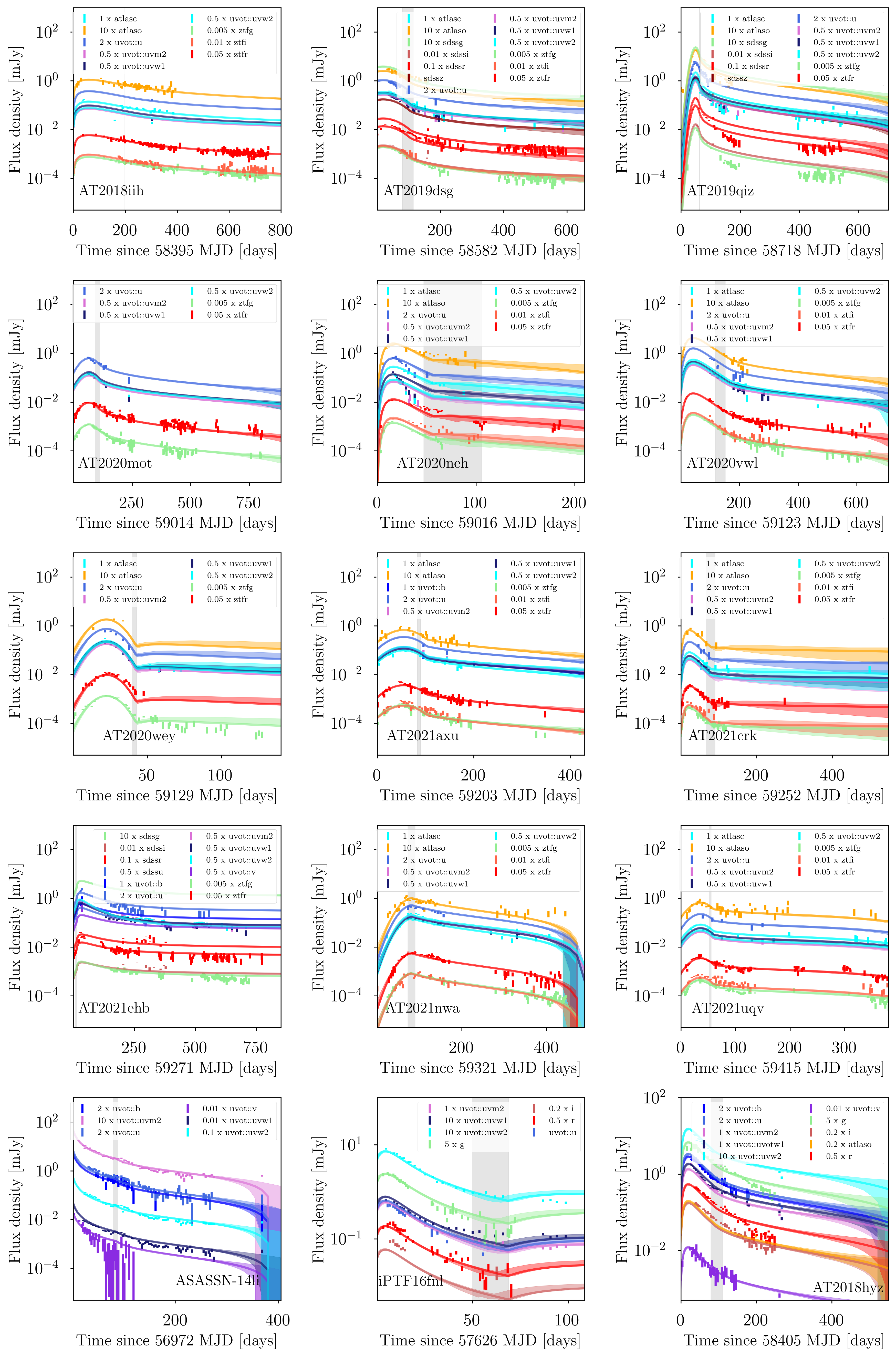} \hspace{-.4cm}
  \end{tabular}
  \caption{Light curves for a subset of TDEs from our sample. The green and red markers indicate the ztfg and ztfr band for different TDEs, while the band represents the $90\%$ credible interval from our fit. The vertical black band indicates the $90\%$ credible interval for $\chi\tfb{}$, i.e., where the fit transitions from the gaussian rise to the cooling envelope.} 
\label{fig:lightcurves}  
\end{figure*}

The cooling envelope model describes the UV/optical light curve only after the debris has circularized into an envelope but not during the circularization/envelope-formation process. The timescale for the latter is uncertain but is likely to be around when the UV/optical light curve peaks (e.g., \citealt{Steinberg&Stone22}) and if circularization is indeed rapid, will be comparable to the fall-back timescale (e.g., \citealt{Stone+13}), 
\begin{equation}
\tfb{} \approx \unit[58]{d} \left(\frac{k}{0.8}\right)^{-3/2} m_{\star}^{1/5} M_{\rm{BH, 6}}^{1/2}.
\label{eq:tfb}
\end{equation} 
Here $k$ is the binding energy constant, which we take to be $0.8$ throughout (corresponding to a $\beta = 1$ encounter for a $\gamma = 5/3$ polytrope; e.g., \citealt{Guillochon&RamirezRuiz13}). We explore the sensitivity of our results to different assumed values of $k$ in Appendix \ref{sec:modelsensitivity}.

However, to fit the complete set of observations, we need a model to capture the behaviour of the light curve before circularization completes and the envelope fully forms. Previous work fitting TDE observations with phenomenological models have modelled the light curve before and near peak to be with a Gaussian or power-law~\citep{vanVelzen+21, yao23}. 
We use a similar approach and model the optical light curve before the time $\chi \tfb$ as either an exponential power law or a Gaussian rise which smoothly connects to the \citetalias{Metzger22} cooling envelope model thereafter. We present results in all subsequent sections using the best fitting (as indicated by the Bayesian evidence) pre-cooling envelope phenomenology. Here $\chi$ is a free parameter which accounts for our ignorance in the number of fall-back times required to form the envelope. With the Gaussian rise parameterization to describe the light curve before $\chi \tfb{}$, our full model for the optical light curve is
\begin{equation}\label{eq.gaussianrisetde}
F(\nu, t) = 
\begin{cases}
A_{\nu} e^{-(t - t_{\rm peak})^2/2\sigma_t^2}, & t < \chi \tfb{} \\
\frac{2 \pi h \nu^{3}}{c^{2}} \frac{1}{\exp \left[h\nu / k_{B} T_{\mathrm{eff}}(t)\right]-1} \frac{R_{\mathrm{ph}}^{2}(t)}{d_{L}^{2}}, & t \geq \chi \tfb{},
\end{cases}
\end{equation}
where $d_{L}$ is the luminosity distance, $h$ is Planck's constant, $k_{B}$ is the Boltzmann constant, and $c$ is the speed of light.  The first term describes a Gaussian rise which continues until \tfb{}, described by three parameters, $t_{\rm peak}$ i.e., the peak timescale of the light curve, $\sigma_t$ which dictates the sharpness of the peak and $A_{\nu}$, a normalization for each specific frequency, $\nu$. The second term describes the emission from the cooling envelope model following \citetalias{Metzger22}. For the exponential power law profile, the lightcurve at times $t < \chi\tfb{}$ instead follows
\begin{equation}
F(\nu, t) = A_{\nu} \left(1 - e^{-t/t_{\rm peak}}\right)^{\alpha_1} \left(\frac{t}{t_{\rm peak}}\right)^{-\alpha_2},
\end{equation}
where and $\alpha_1$ and $\alpha_2$ are the power law slopes for the rise and decay post-peak respectively. We note that for either phenomenological model, $A_{\nu}$ is not used in fitting and is set by enforcing the phenomenological model smoothly connects with the cooling envelope model. 

As summarized in Table \ref{tab:modelparams} in Appendix \ref{sec:scalings}, the full model (cooling envelope and gaussian rise) can be described by $8$ parameters: the $2$ ($3$) parameters that describe the gaussian rise (exponential power law) mentioned above; $\chi$ which relates the time the lightcurve transitions to the cooling envelope as a multiple of the fall-back timescale, the SMBH and star masses ($M_{\rm{BH}}$, $M_{\star}$); the feedback efficiency, $\eta$, with which SMBH accretion adds energy to the envelope; the \citet{Shakura&Sunyaev73} viscosity parameter of the accretion disk, $\alpha$; and the orbital penetration factor, $\beta.$  The other model parameters from \citetalias{Metzger22}, the stream penetration factor $\zeta = 2$ and disk aspect ratio $H/r = 0.3$, are fixed at their fiducial values. 

\section{Results} \label{sec:results}
\begin{figure}[ht!]
\centering
\includegraphics[width=0.45\textwidth]{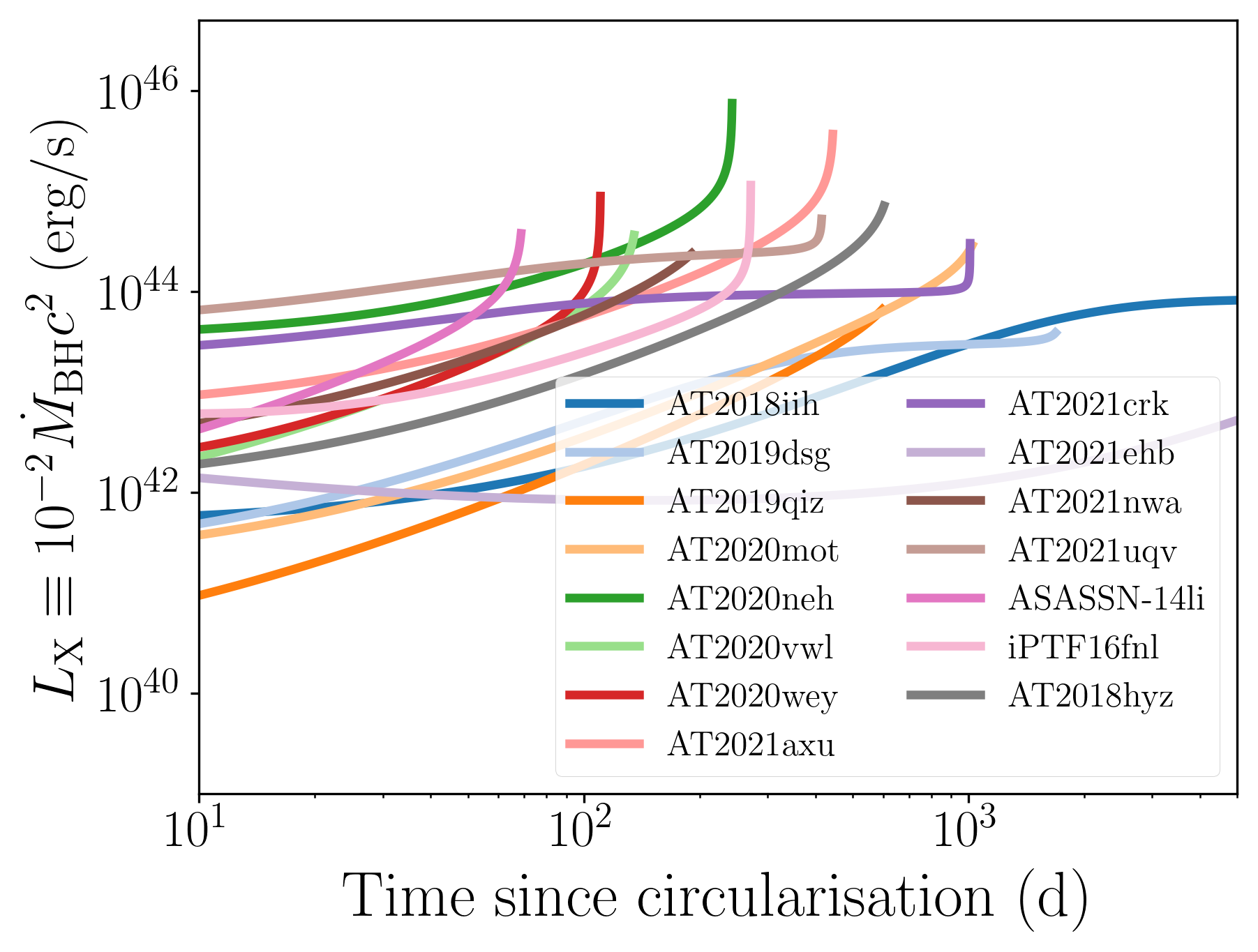} \hspace{-.4cm}
\caption{Proxy X-ray light curves $L_{\rm X} \equiv 10^{-2}\dot{M}_{\rm BH}c^{2}$ of the best-fit models for all TDEs in our sample, where $\dot{M}_{\rm BH}(t)$ is the SMBH accretion rate.  Insofar that X-ray emission from the inner accretion disk may only be visible through the initially narrow polar funnel of the envelope, these luminosities may reflect those observed for a fraction of viewing angles.} 
\label{fig:xray}  
\end{figure}

\begin{figure*}[ht!]
\centering
\includegraphics[width=0.95\textwidth]{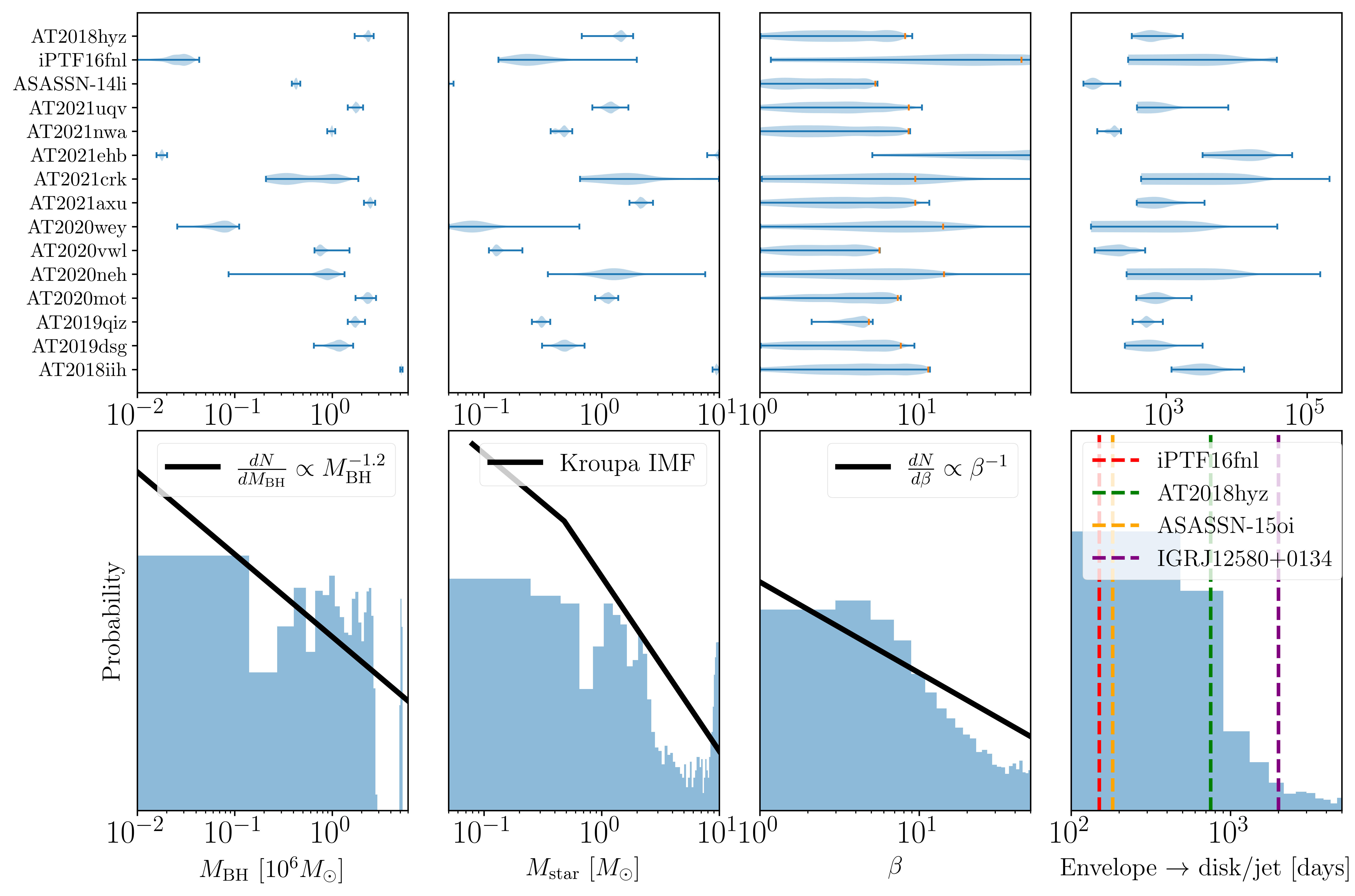} \hspace{-.4cm}
\caption{Violin plots showing (from left to right) the inferred black hole mass, disrupted star mass, TDE penetration factor $\beta \equiv r_{\rm p}/r_{\rm t}$, and the time since optical peak when the envelope has fully contracted to the circularization radius ($R_{\rm v} = R_{\rm c}$) and the SMBH accretion rate peaks.  The bottom panel shows event-averaged distributions for each parameter (in blue) alongside the expectations of TDE loss-cone analysis (black lines; see text for details). The orange ticks in the third panel show the maximum penetration factor $\beta_{\rm max}(M_{\star},M_{\rm BH})$ imposed by the Hills limit given the most probable SMBH and star mass for each event. For comparison we also show in the final panel the observed timescales relative to the optical peak of delayed radio transients, nominally associated in the cooling envelope with when the SMBH accretion rate peaks (References: iPTF 16fnl \citep{Horesh+21}; AT2018hyz \citep{Cendes+22}; ASASSN-15oi \citep{Horesh+21b}; IGRJ12580+0134 \citep{Perlman+22}).} 
\label{fig:violin}  
\end{figure*}

We now confront observations with the model introduced above. We employ a largely homogeneous sample of TDEs discovered with ZTF~\citep{vanVelzen+21, yao23} to minimize the influence of potential selection effects. We also include three additional notable TDEs from the literature~\citep{Holoien2016, Blagorodnova+17, Holoien+20}. We only include events in our sample with well-observed light curves near peak to best constrain the transition point between the pre-cooling envelope phenomenology and cooling-envelope model itself, ignoring TDEs that suggest an optical rebrightening as such behavior is likely due to physical processes not captured by the cooling envelope model. 
We further restrict our sample to only include TDEs from galaxies with measured bulge velocity dispersion $\sigma_{\rm gal}$ to allow comparison between our estimated SMBH masses and those predicted by the $M_{\rm BH}-\sigma_{\rm gal}$ relation. Our selected sample of TDEs is summarised in Table.~\ref{tab:sample}.

\begin{table}
\centering
\caption{Sample of TDEs analyzed in this work alongside the redshift, $\sigma_{\rm gal}$ measurement, and TDE spectral type. The data for our sub sample is available as part of the larger sample collated in~\citet{yao23} and references therein or the open access catalog \citep{oac}.}
\begin{tabular}{llrr}
\toprule
Event & Redshift & $\sigma_{\rm gal}$ [km s$^{-1}$] & Spectral Subtype\\
\midrule
AT2018iih & 0.21 & $148.64\pm 14.42$ & TDE-He \\
AT2019dsg & 0.05 & $86.89\pm 3.92$ & TDE-H+He  \\
AT2019qiz & 0.02 & $69.70\pm 2.30$  & TDE-H+He  \\
AT2020mot & 0.07 & $76.61\pm 5.33$ & TDE-H+He  \\
AT2020neh & 0.06 & $40.00\pm 6.00$ & TDE-H+He  \\
AT2020vwl & 0.03 & $48.49\pm 2.00$ & TDE-H+He  \\
AT2020wey & 0.03 & $39.36\pm 2.79$ & TDE-H+He  \\
AT2021axu & 0.19 & $73.50\pm 17.56$ & TDE-H+He  \\
AT2021crk & 0.15 & $57.62\pm6.29$ & TDE-H+He  \\
AT2021ehb & 0.02 & $99.58\pm3.83$ & TDE-featureless  \\
AT2021nwa & 0.05 & $102.44\pm5.37$ & TDE-H+He  \\
AT2021uqv & 0.11 & $62.30\pm7.08$ & TDE-H+He  \\
ASASSN-14li & 0.0205 & $81\pm2$ & TDE-Bowen \\
IPTF16fnl & 0.0163 & $89\pm1$ & TDE-Bowen \\
AT2018hyz & 0.0458 & $60\pm5$ & TDE-H \\
\bottomrule
\label{tab:sample}
\end{tabular}
\end{table}

For each of the TDEs in our sample we fit the multi-band photometry (detected at greater than $3\sigma$) from \citet{yao23} with the model described above using the open-source software package \program{Redback}~\citep{sarin23_redback}, with the \program{dynesty} sampler~\citep{speagle20} implemented in \program{Bilby}~\citep{Ashton2019a, RomeroShaw2020a}. We sample with a Gaussian likelihood and an additional white noise term. 
We use broad uniform priors on all parameters mentioned above and also sample in the unknown start time of the TDE light curve, with a uniform prior of up to 50 days before the first observation and an additional extinction term, $A_{v}$. 
To ease sampling, we additionally place two constraints on our priors that $\beta \lesssim \beta_{\rm max}$, where $\beta_{\rm max}$ is the maximum penetration factor corresponding to the Hills limit for a non-spinning SMBH (see \citetalias{Metzger22} for details) and $\chi\tfb{}$ is within $\unit[200]{d}$ of the optical lightcurve peak. For the exponential rise parameterisation, we also constrain the priors to obey $\alpha_1 \geq \alpha_2$, i.e., to enforce a rise then decay light curve shape.

Fig.~\ref{fig:lightcurves} shows the fits to all TDE light curves from our sample for the best fit phenomenology, demonstrating that the model can explain the data well. In Appendix \ref{sec:supplementary}, we present a representative corner plot of AT2020mot highlighting the covariances between the parameters characteristic of all our fits.
Across our sample, we see a wide range of light curve rise timescales, which are adequately explained by the pre-cooling envelope phenomenology with the cooling envelope model providing a good fit to the data after $\chi\tfb{}$ (shaded in black). Fig.~\ref{fig:xray} shows the predicted proxy X-ray light curves from the best-fit model for each TDE in our sample, which we assume to track the time evolution of the SMBH accretion rate predicted by the model. The X-ray lightcurves have been truncated at the Eddington luminosity of the SMBH or when the envelope evolution terminates, whichever occurs first. 

In Fig.~\ref{fig:violin}, we show violin plots for all TDEs analyzed in this work of the inferred SMBH mass, the mass of the disrupted star, the orbital penetration factor, $\beta$, and the envelope contraction time (i.e., the timescale for the envelope to contract to the circularization radius and the SMBH accretion rate to peak). We also plot in the panel below histograms of the inferred parameter distributions (equal-weighted by each TDE in the sample) alongside theoretically motivated distributions from TDE loss-cone theory, as we discuss in the next section. The two other parameters in the cooling envelope model; the feedback efficiency, $\eta$ and the disk viscosity, $\alpha$ are largely unconstrained from the priors for our entire sample. In Appendix~\ref{sec:supplementary}, we provide the constraints on several parameters and explore the sensitivity of these results to the modelling assumptions.

\section{Discussion}
\label{sec:discussion}

\subsection{SMBH and Stellar Demographics}

The application of loss-cone theory to a large sample of galactic nuclei predicts the distribution of TDE with SMBH mass of the form $dN/dM_{\rm BH} \propto M_{\rm BH}^{-\delta}$, where $\delta \approx 1.2$ for galaxies with cusp-like nuclei which dominate the TDE rates by the low-mass $\sim 10^{6}M_{\odot}$ SMBH of interest (\citealt{Stone&Metzger16}; see also \citealt{Magorrian&Tremaine99,Wang&Merritt04,Kochanek16a}). Our black hole distribution shown in the left panel of Fig.~\ref{fig:violin} is roughly consistent with this expectation but is underrepresented at the lowest black hole masses $M_{\rm BH} \lesssim 5\times 10^{5}M_{\odot}$. This deficit may in part be a selection effect: in Eddington-limited models (including the cooling envelope scenario), low-$M_{\rm BH}$ TDEs are less luminous and hence would be under-counted in a flux-limited survey such as ZTF. Also note that while TDE rate estimates such as those in \citet{Metzger&Stone16} assume that all galactic nuclei contain SMBH, the SMBH occupation fraction is not well constrained in this mass range (e.g., \citealt{Greene&Ho07}) and may well decrease moving towards lower SMBH mass.  

Our sample also exhibits a less pronounced deficit of massive SMBH with $M_{\rm BH} \gtrsim 10^{6.5}M_{\odot}$ compared to expectations. Again, we speculate that this deficit could be a product of Malmquist bias (brighter TDE from more massive BH) and the selection effect that the peak duration of the optically-luminous phase in the cooling envelope model scales as $t_{\rm KH} \propto M_{\rm BH}^{-2/3}$ (Eq.~\eqref{eq:tKH}) and hence the transition to an optically-dim (but X-ray luminous) disk phase will be more rapid for higher $M_{\rm BH},$ leading to them being missed in optical transient surveys (though possibly contributing to ``X-ray only'' TDEs such as those which dominate the {\it eROSITA} sample; \citealt{Sazonov+21}). In agreement with our findings, \citet{Wevers+19} found that the majority of optically-selected TDEs arise from SMBH masses centered around $\sim 10^{6}M_{\odot}$. In contrast, a comparatively greater fraction of X-ray selected TDEs arise from higher-mass SMBH. However, 
\citet{Wevers+19} also identify a population of X-ray TDEs in significantly less massive galaxies potentially hosting very low-mass black holes $\sim 10^{5}M_{\odot}$. In the cooling envelope scenario, such events may be explained by the disruption of very low-mass stars or brown dwarfs, for which the envelope cooling timescale $t_{\rm KH} \propto m_{\star}^{13/15}$ (Eq.~\eqref{eq:tKH}) is again short despite the low SMBH mass as a result of the low envelope mass.

\begin{figure}[ht!]
\centering
\includegraphics[width=0.45\textwidth]{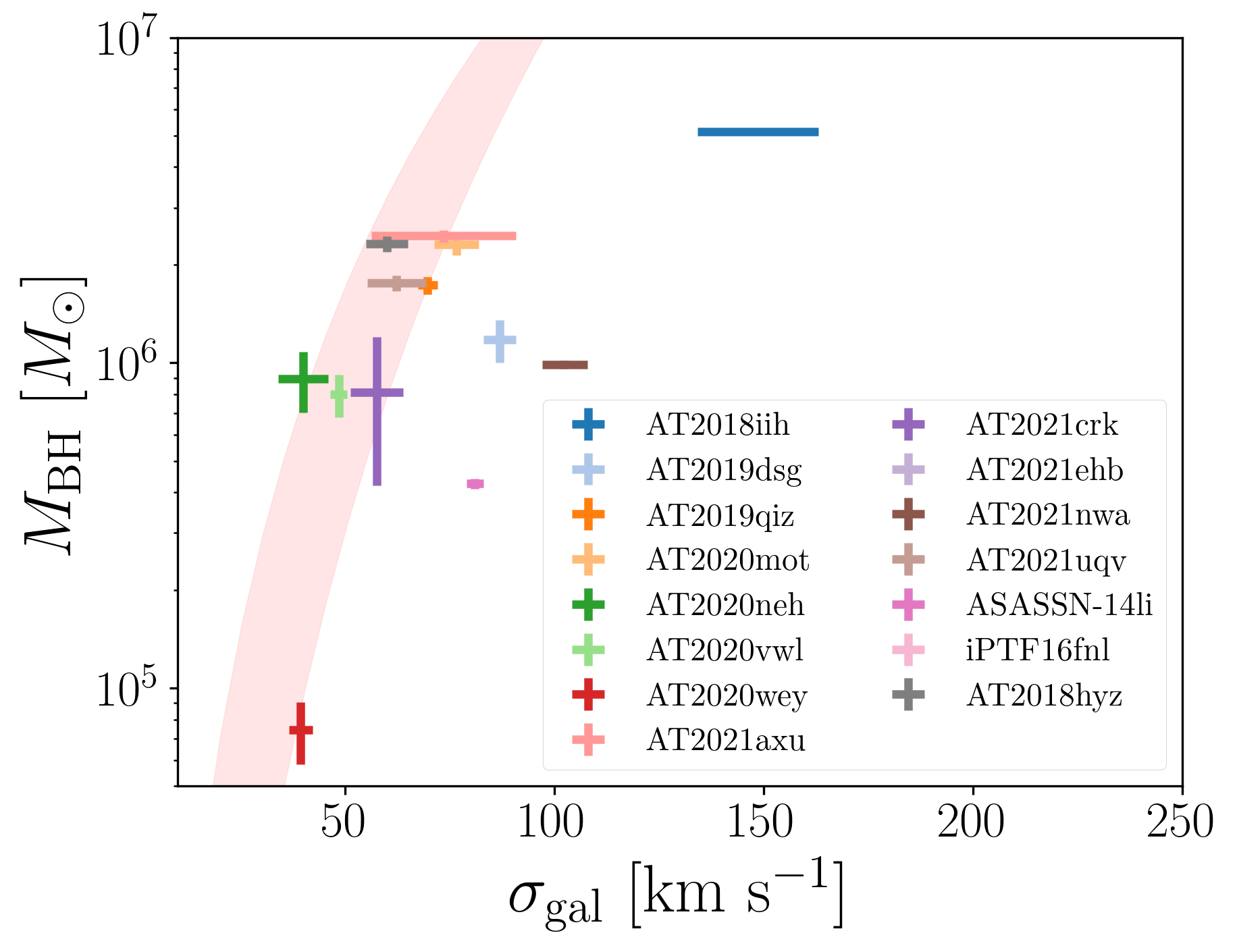} \hspace{-.4cm}
\caption{Inferred black hole masses from the application of our model versus the observed galaxy bulge velocity dispersion $\sigma_{\rm gal}$. The red band shows the $3\sigma$ contour $M_{\rm BH}-\sigma_{\rm gal}$ relation from \citet{Kormendy2013}.} 
\label{fig:mvssigma}  
\end{figure}

A related consideration is how our black hole masses compare to those predicted by the empirical $M_{\rm BH}-\sigma_{\rm gal}$ relation (e.g., \citealt{Gultekin2009}). Fig.~\ref{fig:mvssigma} shows our constraints on $M_{\rm BH}$ alongside the measured $\sigma_{\rm gal}$ from Table~\ref{tab:sample}.  While most of our sample is consistent with the relation (red-shaded band), there are a handful of exceptions. Such inconsistencies, which has also been seen in a fraction of TDEs in previous analyses~\citep{Mockler+19, Ryu+20}, may indicate different physical processes at work in such events leading to inaccurate mass measurement or point towards a failure of the empirical $M_{\rm BH}-\sigma_{\rm gal}$ relation in this SMBH mass range. 
Another selection effect that may lead to outliers such as AT2018iih is that the Hills criterion limits stellar disruptions in massive galaxies to only those hosting SMBH with anomalously low masses \citep{Ramsden+22}.

In Fig.~\ref{fig:violin} we compare our inferred distribution of stellar masses to the \citet{Kroupa01} initial mass function\footnote{In detail, the disruption rate distribution $dr/dM_{\star}$ follows the IMF $dN/dM_{\star}$ according to $dr/dM_{\star} \propto (R_{\star}^{1/4}/M_{\star}^{1/12})(dN/dM_{\star}) \propto M_{\star}^{7/60}(dN/dM_{\star})$ (e.g., \citealt{MacLeod+12,Kochanek16a}) where in the final line we have used $R_{\star} \propto M_{\star}^{4/5}.$  However, this correction factor $M_{\star}^{7/60}$ is a weak dependence relative to uncertainties on the IMF itself.} (IMF), finding reasonable agreement at the high-$M_{\star}$ end but a potential deficit of low-mass stars $M_{\star} \lesssim 0.5M_{\odot}.$  
Again, we speculate that this could be a selection effect insofar that $t_{\rm KH} \propto M_{\star}^{13/15}$ (Eq.~\eqref{eq:tKH}) and hence the optically-luminous phase could be considerably shorter for low-mass stars. \citet{Nicholl+22} found a similar deficit of low-$M_{\star}$ events, which they attribute to the smaller range of SMBH masses capable of disrupting low-mass stars due to their smaller radii. 
\citet{Mockler+22} found that massive stars ($\gtrsim 1-2M_{\odot})$ are over-represented among the TDE sample by a factor of $\gtrsim 10^{2}$ to account for the high nitrogen-to-carbon abundances they infer from UV line ratios (see also \citealt{Kochanek16b}). Normalized over the entire mass range of the IMF, we also find $M_{\star} > 1 M_{\odot}$ stars to be at least moderately over-represented in our sample. However, this must be caveated by our small sample size (dominated by the high inferred $M_{\star}$ of AT2018iih).

Most TDEs by SMBH of mass $M_{\rm BH} \sim 10^{6}M_{\odot}$, are predicted to occur in the so-called ``pinhole'' (or full loss-cone) regime, in which stars wander in and out of the loss-cone several times in its final orbit before being tidally disrupted (e.g., \citealt{Stone&Metzger16}, their Eq.~29). The penetration factor distribution in the pinhole regime is predicted to obey $dN/d\beta \propto \beta^{-1}$, for values of $\beta$ between the disruption limit ($\beta \approx 1-2$) and the maximum value $\beta_{\rm max} \simeq 12\, m_{\star}^{7/15}M_{\rm BH,6}^{-2/3}$ corresponding to the Hills limit\footnote{Even in the non-pinhole (``diffusive'') regime, a significant fraction of TDEs can occur with $\beta \gg 1$ (e.g., \citealt{Weissbein&Sari21}).}.
The inferred distribution of penetration factors from our TDE sample is somewhat steeper than $\propto 1/\beta$ for $\beta \gtrsim 10$. However, this is a consequence of our imposed constraint that $\beta \lesssim \beta_{\rm max}$, i.e., that the penetration factor is less than the Hills limit for a non-spinning SMBH combined with the constraint that $\chi\tfb{} \lesssim \unit[200]{d}$. In Fig.~\ref{fig:violin}, we indicate (with orange ticks) the Hills limit corresponding to the most probable black hole and stellar mass for each TDE. For most TDEs in our sample, our constraints on $\beta$ are effectively the prior on $\beta$ conditioned on the inferred masses, resulting in the steeper than $\propto 1/\beta$ beyond $\beta \approx 10$ (comparable to the average Hills limit of our sample). 

\subsection{X-ray/Radio Emission from Delayed SMBH Feeding}

Several TDE observations hint that the peak of the SMBH accretion rate is significantly delayed with respect to when the optical light curve peaks. While a handful of powerful jetted TDEs exhibit bright non-thermal X-ray and radio emission (e.g., \citealt{Bloom+11,Burrows+11}), most TDEs are radio dim (e.g., \citealt{Alexander+20}) excluding powerful off-axis jets (e.g., \citealt{Generozov+17}). Nevertheless, several TDEs exhibit late-time radio flares, indicating mildly relativistic material ejected from the vicinity of the SMBH but delayed from the optical peak by several months to years (e.g., \citealt{Horesh+21,Horesh+21b,Sfaradi+22,Cendes+22,Perlman+22}).  A potentially related occurrence is the coincident detection of high-energy neutrinos from several optical TDEs by IceCube \citep{Stein+21,vanVelzen+21b,Reusch+22}, each of which also arrived several months after the optical peak.

The final panel of Fig.~\ref{fig:violin} constrains the envelope termination time of our best-fit models, defined as the number of days after the optical peak at which the envelope has contracted to the circularization radius and hence any remaining envelope mass has collapsed into a centrifugally-supported disk. Around this same time, the accretion rate onto the SMBH reaches its peak (Fig.~\ref{fig:xray}), potentially leading to sudden onset of powerful disk outflows or a relativistic jet and associated synchrotron radio emission \citep{Giannios&Metzger11}.  For most of our sample, the envelope termination timescale is poorly constrained. However, the event-averaged allowed range overlaps with the timescales of delayed radio emission seen in previous TDEs~\citep{Horesh+21,Horesh+21b,Cendes+22,Perlman+22}. 

A few of the TDEs in our sample, e.g., AT2018hyz, ASASSN-14li, AT2021nwa, AT2019qiz, and AT2019dsg have well measured envelope termination times, all of which are around $\approx \unit[1000]{d}$ after optical peak. Two of these events have published radio detections; AT2020vwl~\citep{Goodwin2023} which was first detected in radio $\approx \unit[100]{days}$ after optical peak and observed for up to $\approx \unit[500]{d}$ after, while AT2019dsg had radio observations from $50-560\, \rm{d}$ post disruption~\citep{Cendes2021}.  The envelope termination timescales we predict for both these TDEs are inconsistent with the early radio detections in the systems; however, the observed radio emission may be dominated by prompt outflows generated during the circularization process, rather than delayed radio emission due to the late-onset SMBH accretion predicted in the cooling envelope model. 

Some TDEs in our sample also have credible intervals consistent with sharp drops in luminosity after the last set of observations. This is largely a consequence of our modelling set up and the implicit assumption that there is no optical/UV emission after envelope contraction. In reality, once the envelope fully contracts to an centrifugally-supported disk (after which our model terminates), UV/optical emission from the disk will take over. \citet{Mummery+24} find that the plateau-shaped late-time optical/UV light curves of many TDEs can be explained as multi-color blackbody thermal emission from a viscously spreading accretion disk. However, because the cooling envelope model can {\it also} generate a flat optical light curve (when envelope cooling/contraction is temporarily balanced by SMBH heating), these two models be challenging to distinguish based on UV/optical data alone.

The proxy X-ray light curves for our best-fit models in Fig.~\ref{fig:xray} demonstrate how the cooling envelope model can also explain the delayed rise of X-ray emission seen in some TDEs (e.g., \citealt{Gezari+17,Yao+22}), at least for those viewing angles which enable X-rays to escape through the envelope's narrow polar funnel.  The similarly delayed establishment of an standard accretion disk is also clear in late-time UV observations (e.g., \citealt{vanVelzen+19}). The SMBH accretion rate reaches its peak value once the envelope's radius contracts to the circularization radius (rightmost panel in Fig.~\ref{fig:violin}); this signals the end of the envelope phase and the beginning of the disk-dominated accretion phase (e.g., \citealt{Shen&Matzner14,Mummery&Balbus20}).

While X-ray limits exist on some TDE in our sample (e.g., AT2020mot, AT2020wey; \citealt{Hammerstein+23}), a few events indeed show delayed X-ray rises. The X-ray light curve of AT2021ehb peaked roughly 260 d after the optical peak, an epoch associated with strong spectral variability \citep{Yao+22}.  While this is somewhat earlier than the $\sim 10^{3}$ d delay our model fits of this event predict (right panel of Fig.~\ref{fig:violin}), this disagreement is not too disconcerting given the crudeness of how the cooling envelope model treats the final stages of the envelope contraction phase (e.g., when neglecting rotational support becomes indefensible). \citet{Nicholl+20} found that the X-ray light curve of AT2019qiz peaked about a month after the optical maximum, somewhat earlier than our posterior prediction for this event of several months. X-ray observations are also available for AT2018hyz and iPTF16fnl; our constraints on the envelope termination timescales are consistent for AT2018hyz but inconsistent for iPTF16fnl. Again, the latter is not disconcerting given uncertainties of how the model treats the final stages of the envelope contraction.

\subsection{Comparison to Other Population Studies}

Five of our TDEs have been previously analyzed with physically motivated models; \texttt{MOSFiT}~\citep{Guillochon+18,Nicholl+22, Hammerstein+23} and \texttt{TDEmass}~\citep{Ryu+20}. The emission processes invoked in these models differ significantly from the cooling envelope model; the former assumes the light curve scales with the mass fall-back rate~\citep{Guillochon+18} while the latter invokes shocks between the debris streams as the source of UV/optical emission~\citep{Piran+15,Ryu+23}. A comparison of the system properties inferred by these models can shed light into systematics, particularly for drawing comparisons to theoretical distributions described above. In Appendix \ref{sec:supplementary}, we compare our inferred black hole and stellar masses to those from \texttt{MOSFiT} and \texttt{TDEmass}. 

The cooling envelope model generally predicts smaller black hole and stellar masses than either \texttt{MOSFiT} or \texttt{TDEmass}. In most cases the discrepancies in black hole masses between the cooling envelope model and \texttt{TDEmass} are small, which is encouraging for the prospects of using TDEs to probe SMBH demographics. By contrast, the black hole masses of both the cooling envelope model and \texttt{TDEmass} exhibit larger differences compared to those from \texttt{MOSFiT}. This discrepancy could hint at limitations of the assumption in \texttt{MOSFiT} that the disrupted material circularizes promptly into an accretion disk or in the phenomenological treatment of the photosphere radius.

Unlike the black hole masses, the discrepancy is considerably larger between the three models for the inferred stellar mass, with both \texttt{MOSFiT} and \texttt{TDEmass} predicting stellar masses an order of magnitude larger than our constraints (though we note that our stellar mass distribution is closer to that found by \citealt{Nicholl+22} using \texttt{MOSFiT} than by \citealt{Hammerstein+23}). This disagreement is particularly stark for AT2018iih, where we constrain the mass to be $\unit[9.5\pm0.2]{M_{\odot}}$ (the largest in our sample) versus $\unit[75^{+14}_{-13}]{M_{\odot}}$ with \texttt{TDEmass}.  For a Kroupa-like IMF, most TDEs should indeed originate from low-mass stars, which may point to an issue for both \texttt{MOSFiT} and \texttt{TDEmass}; however, as already discussed, there is evidence that massive stars may be over-represented in the TDE sample (\citealt{Kochanek16b,Mockler+22}).  The current sample of events is too small to draw any strong conclusions, particularly in light of the many selection effects discussed above that also differ between the models. Part of the reason for our low-inferred stellar masses in that the cooling envelope has a high radiative efficiency: an order-unity fraction of the gravitational binding energy of a Keplerian disk at the circularization radius is released as the envelope contracts from its weakly bound initial state to form the disk, allowing a lower envelope (disrupted star) mass to generate a given radiated energy.  

TDE exhibit several spectroscopic classes: TDE-H, TDE-He, and TDE-H+He, depending on whether they exhibit almost exclusively HI lines, a small number of only He II lines, or a mixture of H I, He II, and N III respectively (e.g., \citealt{Gezari+12,Arcavi+14,Holoien+16,vanVelzen+21}).  \citet{Nicholl+22} found that (1) TDE-H arrive on average from less massive SMBH than TDE-He or TDE-H+He classes; (2) a larger fraction of TDE-He/H+He exhibit X-ray emission than TDE-H (see also \citealt{Hammerstein+23}); (3) the ``efficiency'' with which fall-back accretion power is converted to observed luminosity increases with SMBH mass. Insofar that all but two TDEs in our sample are within the TDE-H+He class (Table~\ref{tab:sample}), we cannot explore systematic trends of the star or SMBH properties with spectroscopic class. The only TDE-He event in our sample (AT2018iih) is indeed found to host the largest SMBH mass. In Appendix \ref{sec:scalings} we show that many of the $M_{\rm BH}-$dependent trends identified by \citet{vanVelzen+19,Nicholl+22,Hammerstein+23}, including the X-ray/optically-luminous dichotomy (see also \citealt{Wevers+19}), are consistent with scaling predictions of the cooling envelope model. As commented earlier, TDE from more massive SMBH exhibit short envelope cooling timescales $\propto t_{\rm KH} \propto M_{\rm BH}^{-2/3}$, which shortens any optically-bright photosphere phase and hastens the onset of the UV/X-ray bright accretion disk phase. 

\subsection{Sources and size of systematic uncertainties}
The cooling envelope model and inferences made from this model presented here, suffer from a number of systematic uncertainties that require further model development or exploration. 
In Appendix~\ref{sec:modelsensitivity}, we explore the systematic uncertainty associated with the choice of pre-cooling envelope phenomenology. We find that provided the data can be smoothly fitted with the pre-cooling phenomenology and the cooling envelope model itself, then the choice of phenomenology is not a large source of systematic error, with posteriors on salient parameters within the statistical uncertainty across the choice of phenomenology. However, in cases where the data is not informative about the transition (observations not considered in this paper), model parameters are influenced by the choice of phenomenology, and caution must be applied interpreting such lightcurves with the cooling envelope model as currently implemented in \program{Redback}.

Throughout this work, we also fixed the binding energy constant $k=0.8$, i.e., appropriate for a $\beta=1$ encounter for a $\gamma = 5/3$ polytrope. This is not a valid assumption for many of the TDEs analysed in this work. In the Appendix, we present results fitting AT2019dsg with $k=0.4$, finding this to lead to a decrease by a factor of two in the inferred black hole mass, a factor two increase in the inferred stellar mass and produce an uninformative measurement of $\beta$. This represents a source of systematic uncertainty that is $\approx$ twice as large as the statistical uncertainty in the fitted parameters. However, specifically for AT2019dsg, the $k=0.8$ provides an overwhelmingly better fit to the data (a Bayes factor of $\sim 10^{17}$ relative to the $k=0.4$ fit) suggesting that the choice of $k$ as a systematic uncertainty can be removed through Bayesian model comparison.

Another source of systematic uncertainty is the treatment of final stages of envelope contraction; while unlikely to affect the estimates of the stellar and black hole mass, this can have a stronger affect on the inferred envelope contraction time. Furthermore, if a distinct emission source (e.g., thermal emission from the viscously spreading accretion disk; e.g., \citealt{Cannizzo+90,Shen&Matzner14,Mummery+24}) takes over when the cooling envelope phase terminates at late times, then a physical model for this additional emission component would help to properly isolate the transition point.

The current sources of systematic uncertainties can be mitigated and therefore motivate further model development. In particular, the inclusion of the stellar mass and $\beta$ dependence on $k$, will remove the dominant uncertainty in black hole and stellar mass measurements from the choice of $k$. Meanwhile, including a more physically motivated lightcurve pre-cooling envelope will minimize the systematics associated with the choice of phenomenology and also allow the broader model to be confronted with a larger sample of TDE.
\section{Conclusions} \label{sec:conclusion}

We have presented results interpreting a sample of 15 optical TDE observed mainly by the Zwicky Transient Facility within the cooling envelope model in order to constrain properties such as the SMBH mass $M_{\rm BH}$ and stellar mass $M_{\star}$. As summarized in Fig.~\ref{fig:violin}, the distributions of these inferred properties for our sample broadly follow the theoretical expectations of loss-cone analysis assuming a standard stellar initial mass function. 
However, we find a deficit of events with $M_{\rm BH} \lesssim 5\times 10^{5}M_{\odot}$, and $M_{\star} \lesssim 0.5M_{\odot}$, which we speculate could result from the reduced detectability of TDEs with these properties. The more moderate deficit of $M_{\rm BH} \gtrsim 3\times 10^{6}M_{\odot}$ systems in our sample reflects the short cooling envelope cooling time $t_{\rm KH} \propto M_{\rm BH}^{-2/3}$ for high-mass SMBH systems, which would instead preferentially give rise to disk-dominated X-ray luminous TDEs. 
Our inferred SMBH masses of our sample are also broadly consistent with the expectations from the empirical $M_{\rm BH}-\sigma_{\rm gal}$ relation, with a few exceptions which may indicate a failure of the empirical relation or that not all TDEs emit optical/UV radiation via the same physical mechanism. 

We also use our model fits to illustrate the predicted range in delay times between the peak of the optical light curve and when the mass accretion rate onto the SMBH reaches its maximum (once the envelope has contracted to the circularization radius) on a timescale of months to years (Figs.~\ref{fig:violin}, \ref{fig:xray}).  These findings are at least qualitatively consistent with delayed-rising X-ray and non-thermal radio flares seen in a growing sample of TDE, as well as the late-time establishment of a UV bright thermal disk emission phase (e.g., \citealt{vanVelzen+19}). However, we note that multiple mechanisms may contribute to UV/X-ray and radio emission other than the delayed SMBH feeding described here, including prompt outflows or shocks during the early circularization phase (e.g., \citealt{Bonnerot&Lu20,Steinberg&Stone22}) and
late-time state transitions in the disk at critical accretion rates (e.g., \citealt{Giannios&Metzger11,Tchekhovskoy+14}).  
Future multiwavelength observations of our sample may help disentangle these possibilities and provide additional support for the cooling envelope model. 
We note that a number of enhancements of are also needed to confront this model with a broader sample of TDEs, including, a better treatment of the binding energy constant and its dependence on the stellar mass and 
$\beta$, including partial TDEs, and more physically motivated pre-cooling envelope phenomenology.


\section{Acknowledgments}
We thank Kate Alexander, Yvette Cendes, Matt Nicholl, Sjoert van Velzen, and the anonymous reviewer for helpful feedback on the manuscript. N.~Sarin is supported by a Nordita Fellowship. Nordita is supported in part by NordForsk. B.~D.~Metzger is supported in part by the NSF (grant AST-2009255). The Flatiron Institute is supported by the Simons Foundation.

\appendix

\section{Analytic Scalings}
\label{sec:scalings}

\begin{deluxetable}{ccc}
\tablecaption{Cooling envelope model priors\label{tab:modelparams}}
\tablewidth{700pt}
\tabletypesize{\scriptsize}
\tablehead{
\colhead{Symbol} & \colhead{Description} & 
\colhead{Range} 
} 
\startdata
$M_{\star} = m_{\star}M_{\odot}$ & Star mass & $[0.05-20] M_{\odot}$ \\
$M_{\rm BH}$ & SMBH mass & $[10^{-2}-20] 10^{6}M_{\odot}$ \\
$\beta \equiv R_{\rm t}/R_{\rm p}$ & Orbit penetration factor & [1-100] \\
$\alpha$ & Viscosity parameter & $[10^{-4} - 0.01]$ \\
$\eta$ & SMBH feedback efficiency  & $[10^{-4} - 0.01]$\\
\enddata
\end{deluxetable}

Here we summarize some basic analytic scalings of the cooling envelope model with the masses of the star and the SMBH (see \citetalias{Metzger22} for details and prefactors), again assuming $R_{\star} \propto M_{\star}^{4/5}$ appropriate to lower-main sequence stars. The model parameters are summarized in Table \ref{tab:modelparams}.  

A star undergoes tidal disruption once its orbital pericenter radius becomes less than the tidal radius,
\begin{eqnarray}
R_{\rm t} &\approx& R_{\star}(M_{\rm BH}/M_{\star})^{1/3} \propto M_{\star}^{7/15}M_{\rm BH}^{1/3}.
\label{eq:Rt}
\end{eqnarray}
Disruption binds roughly half the star to the SMBH by a specific energy $|E_{\rm t}| \sim GM_{\rm BH}R_{\star}/R_{\rm t}^{2}$. The most tightly bound debris falls back to the SMBH on the characteristic fall-back timescale $t_{\rm fb} \propto M_{\star}^{1/5}M_{\rm BH}^{1/2}$  (Eq.~\eqref{eq:tfb}) set by the period of an orbit with energy $E_{\rm t}$, resulting in a peak mass fall-back rate 
\be
\dot{M}_{\rm fb} \propto M_{\star}/t_{\rm fb} \propto M_{\star}^{4/5}M_{\rm BH}^{-1/2}.
\ee

Assuming circularization to be rapid (e.g., \citealt{Steinberg&Stone22}), the initial radius of the envelope $R_{v,0}$ is found by equating $|E_{\rm t}|$ to half its gravitational binding energy $|E_{\rm b}| \sim GM_{\rm BH}M_{e}/R_{v}$, giving
\begin{eqnarray}
R_{v,0} \sim R_{\star}\left(\frac{M_{\rm BH}}{M_{\star}}\right)^{2/3}\frac{M_{e}}{M_{\star}} \propto M_{\star}^{2/15}M_{\rm BH}^{2/3},
\label{eq:Rv0}
\end{eqnarray}
where we have taken the initial mass of the envelope $M_e$ to be a fixed fraction of $M_{\star}.$

The luminosity of a radiation-supported envelope roughly equals the Eddington luminosity,
\be L_{\rm rad} \approx L_{\rm Edd} \propto M_{\rm BH}.
\ee
The photosphere radius scales with the envelope radius, rendering the effective temperature (Eq.~\eqref{eq:Teff}),
\be
T_{\rm eff} \propto \left(\frac{L_{\rm rad}}{4\pi \sigma R_{v,0}^{2}}\right)^{1/4} \propto M_{\star}^{-1/15}M_{\rm BH}^{-1/12},
\ee
a weak function of the parameters.

The radiated energy at early times in the cooling envelope model is not powered by accretion onto the SMBH.  Nevertheless, if one were to define a ``radiative efficiency'' relative to the fallback accretion power $L_{\rm acc} \propto \dot{M}_{\rm fb}c^{2}$, 
\be
\epsilon \equiv \frac{L_{\rm rad}}{L_{\rm acc}} \propto M_{\star}^{-4/5}M_{\rm BH}^{3/2}.
\ee
After forming, the envelope initially contracts on the Kelvin-Helmholtz timescale $t_{\rm KH}$ (Eq.~\eqref{eq:tKH}) according to
\be
R_{v} = R_{v,0}\left(\frac{t}{t_{\rm KH}}\right)^{-1}.
\ee
Absent feedback from the SMBH, the envelope will therefore contract to the circularization radius ($R_{v} = R_{\rm c} = 2R_{\rm t}/\beta)$, entering the disk-dominated accretion phase, on the timescale,
\be
t_{\rm disk} \propto \left(\frac{R_{v,0}}{R_{\rm t}}\right)t_{\rm KH} \propto
M_{\star}^{8/15}M_{\rm BH}^{-1/3}.
\ee

The true termination time of the envelope (rightmost panel of Fig.~\ref{fig:violin}) can exceed the passive cooling case due to energy input from the SMBH.  

Insofar that the mass fall-back rate obeys $\dot{M} \propto \dot{M}_{\rm fb}(t/t_{\rm fb})^{-5/3}$ at times $t \gg t_{\rm fb},$ the UV/X-ray luminosity achieved after a disk has fully formed at $t \approx t_{\rm disk},$ will scale as 
\be
L_{\rm X}(t_{\rm disk}) \propto \dot{M}(t_{\rm disk}) \propto M_{\star}^{11/45}M_{\rm BH}^{8/9}.
\ee
In summary, the cooling envelope model predicts that optical TDE flares from more massive SMBH should be (1) more luminous $L_{\rm rad} \propto M_{\rm BH}$; (2) exhibit higher ``radiative efficiency'' $\epsilon \propto M_{\rm BH}^{3/2}$ (if normalized to the fall-back accretion power); (3) generate shorter-lived optical flares $\propto t_{\rm KH} \propto M_{\rm BH}^{-2/3}$, which in part therefore (4) produce more luminous UV/X-ray emission $L_{\rm X}(t_{\rm disk}) \propto M_{\rm BH}^{8/9}$ once the envelope contracts and a disk fully forms. These predictions are broadly consistent with the trends in TDE properties with SMBH mass found by \citet{vanVelzen+19,Nicholl+22,Hammerstein+23}.

\section{Parameter Constraints}\label{sec:supplementary}

\begin{figure*}[ht!]
\centering
\includegraphics[width=0.95\textwidth]{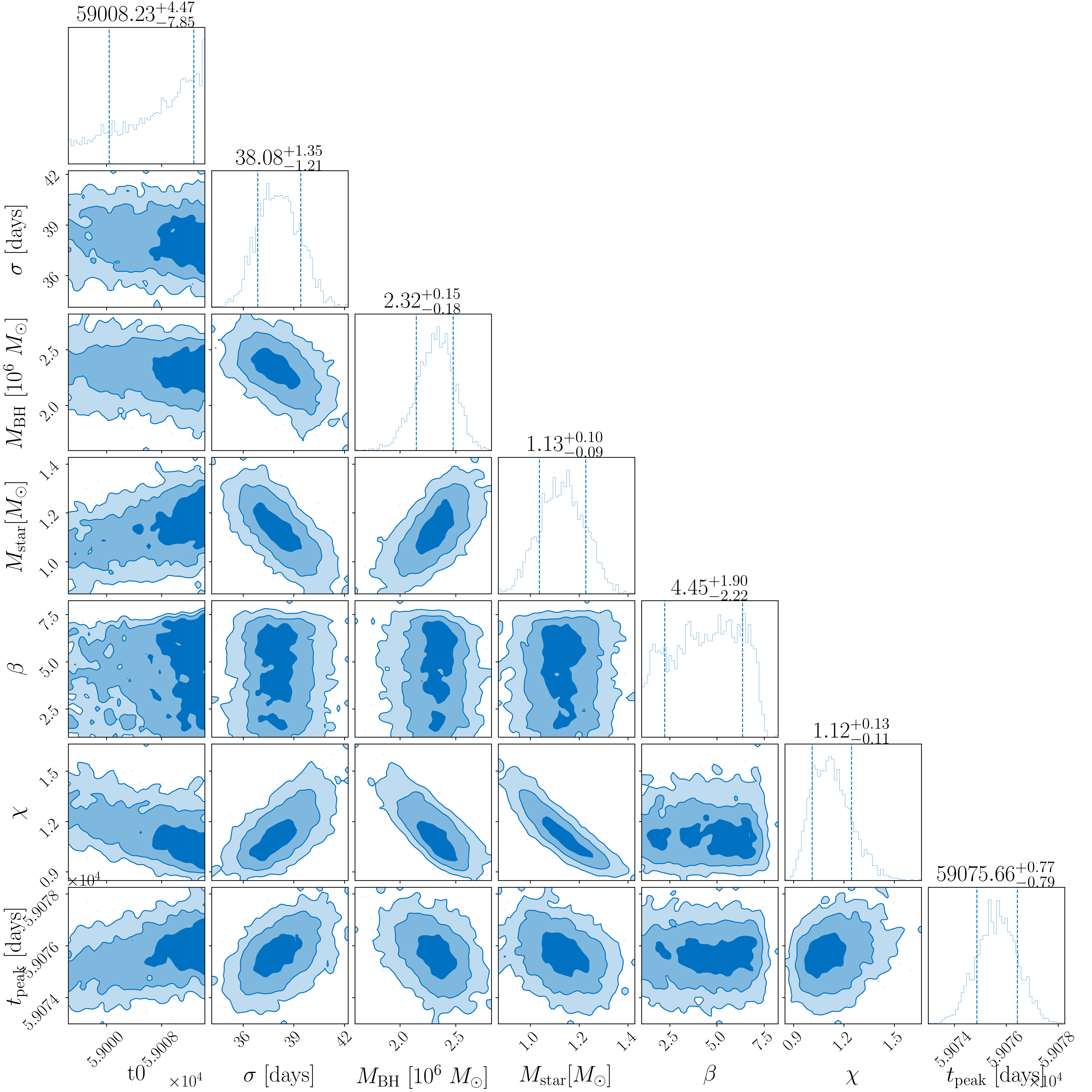} \hspace{-.4cm}
\caption{Example corner plot showing a subset of the parameters from our analysis on AT2020mot.} 
\label{fig:corner}  
\end{figure*}

Fig.~\ref{fig:corner} shows the corner plot from fitting AT2020mot, demonstrating the typical covariances between different model parameters consistent with all other TDEs analyzed in this work. 

In Table~\ref{tab:constraints}, we show the $1\sigma$ constraints on well constrained model parameters for all the TDEs analysed in this work alongside the Bayesian evidence. Table~\ref{tab:comparison} compares the inferred black hole and stellar masses for the TDEs in our sample with previously published fits using \program{MOSFiT} and \program{TDEmass}.

\begin{table*}[]
\centering
\caption{$1\sigma$ constraints on salient parameters for all TDEs analysed in this work alongside the Bayesian evidence $\ln Z$ to indicate the quality of the fit as well as the preferred pre-cooling envelope phenomenology.}
\label{tab:constraints}
\begin{tabular}{@{}llllllll@{}}
\toprule
 & $M_{\rm BH}$ {[}$10^6 M_{\odot}${]} & $M_{\star}$ {[}$M_{\odot}${]} & $\beta$ & $\chi$ & Envelope $\to$ disk/jet [MJD] & $\ln Z$ & Preferred model \\ \midrule
AT2018iih & ${5.13} \pm 0.06 $ &  ${9.49} \pm 0.25 $& ${6.26}_{-3.30}^{+3.42}$ & ${0.801} \pm 0.001$ & ${62259.53}_{-1421.12}^{+3464.65}$ & $2898.9$ &  Power law \\
AT2019dsg &  ${1.18}_{-0.19}^{+0.18}$ &  ${0.49}_{-0.06}^{+0.07}$& ${4.59}_{-2.33}^{+2.12}$ & ${1.85}_{-0.29}^{+0.39}$ & ${59259.24}_{-236.46}^{+420.54}$ & $700.8$ &  Gaussian rise \\
AT2019qiz &  ${1.73}_{-0.10}^{+0.11}$ & ${0.31} \pm 0.02 $ &  ${4.08}_{-0.74}^{+0.52}$& ${1.06}_{-0.06}^{+0.08}$ & ${59326.62}_{-72.62}^{+76.11}$ & $29.0$ &  Gaussian rise \\
AT2020mot &  ${2.32}_{-0.18}^{+0.15}$ &  ${1.13}_{-0.09}^{+0.10}$& ${4.45}_{-2.22}^{+1.90}$ & ${1.12}_{-0.11}^{+0.13}$ &${59959.21}_{-278.04}^{+368.10}$ &  $937.8$ &  Gaussian rise \\
AT2020neh &  ${0.90} \pm 0.17 $&  ${1.39}_{-0.38}^{+0.50}$&  ${7.97}_{-4.68}^{+5.13}$ & ${0.85}_{-0.04}^{+0.12}$& ${60492.14}_{-946.98}^{+3791.27}$ & $287.3$ &  Power law \\
AT2020vwl &  ${0.77}_{-0.05}^{+0.08}$ &  ${0.13} \pm 0.01 $ &  ${3.29}_{-1.52}^{+1.36}$ & ${4.49}_{-0.46}^{+0.32}$ & ${59372.22}_{-72.14}^{+143.59}$ & $772.5$ &  Power law \\
AT2020wey &  ${0.08} \pm 0.02 $ &  ${0.09}_{-0.02}^{+0.05}$ &  ${11.22}_{-6.43}^{+8.31}$ & ${4.58}_{-0.39}^{+0.28}$ & ${60017.35}_{-610.07}^{+2265.11}$ & $182.4$ &  Gaussian rise \\
AT2021axu &  ${2.46}_{-0.11}^{+0.10}$ & ${2.18}_{-0.14}^{+0.15}$ &  ${5.23}_{-2.86}^{+2.75}$ & ${0.81}_{-0.01}^{+0.01}$ & ${60068.70}_{-345.32}^{+777.88}$ &  $933.8$ &  Gaussian rise \\
AT2021crk &  ${0.85}_{-0.52}^{+0.39}$ & ${2.07}_{-0.73}^{+4.72}$ & ${10.16}_{-5.95}^{+19.91}$ & ${1.08}_{-0.16}^{+0.45}$ & ${63425.91}_{-2961.76}^{+15060.34}$ &  $476.2$ &  Power law \\
AT2021ehb &  ${0.02}\pm 0.01$ &  ${9.71}_{-0.42}^{+0.20}$ &  ${8.89}_{-4.88}^{+4.77}$ & ${1.23}_{-0.04}^{+0.04}$ & ${77605.70}_{-8704.16}^{+12685.86}$ &  $665.5$ &  Gaussian rise \\
AT2021nwa &  ${0.99}_{-0.03}^{+0.02}$ &  ${0.47}_{-0.06}^{+0.03}$&  ${4.34}_{-2.54}^{+2.65}$ & ${1.53}_{-0.08}^{+0.28}$ & ${59593.52}_{-33.43}^{+21.33}$ & $1128.8$ &  Gaussian rise \\
AT2021uqv &  ${1.76}_{-0.09}^{+0.10}$&  ${1.20}_{-0.12}^{+0.12}$&  ${4.41}_{-2.27}^{+2.77}$ & ${0.806}_{-0.004}^{+0.009}$& ${59904.21}_{-40.25}^{+94.11}$ & $1127.6$ &  Gaussian rise \\
ASASSN-14li &  ${0.43}_{-0.02}^{+0.01}$ &  ${0.05}_{-0.001}^{+0.001}$ & ${2.93}_{-1.47}^{+1.57}$ & ${4.24}_{-0.13}^{+0.13}$ & ${57055.96}_{-9.28}^{+14.24}$ & $448.3$ &  Power law \\
iPTF16fnl & ${0.03} \pm 0.01$ &  ${0.28}_{-0.09}^{+0.19}$ & ${36.81}_{-22.51}^{+27.50}$ & ${9.23}_{-0.81}^{+0.53}$ & ${62279.92}_{-3575.12}^{+24208.35}$ & $249.0$ &  Power law \\
AT2018hyz & ${2.33}_{-0.14}^{+0.11}$ & ${1.45}_{-0.11}^{+0.09}$ & ${4.34}_{-2.35}^{+2.53}$ & ${0.84}_{-0.03}^{+0.08}$ & ${59243.67}_{-276.26}^{+488.54}$ &  $390.7$ & Power law \\ \bottomrule
\end{tabular}
\end{table*}

\begin{figure*}[ht!]
\centering
\includegraphics[width=0.95\textwidth]{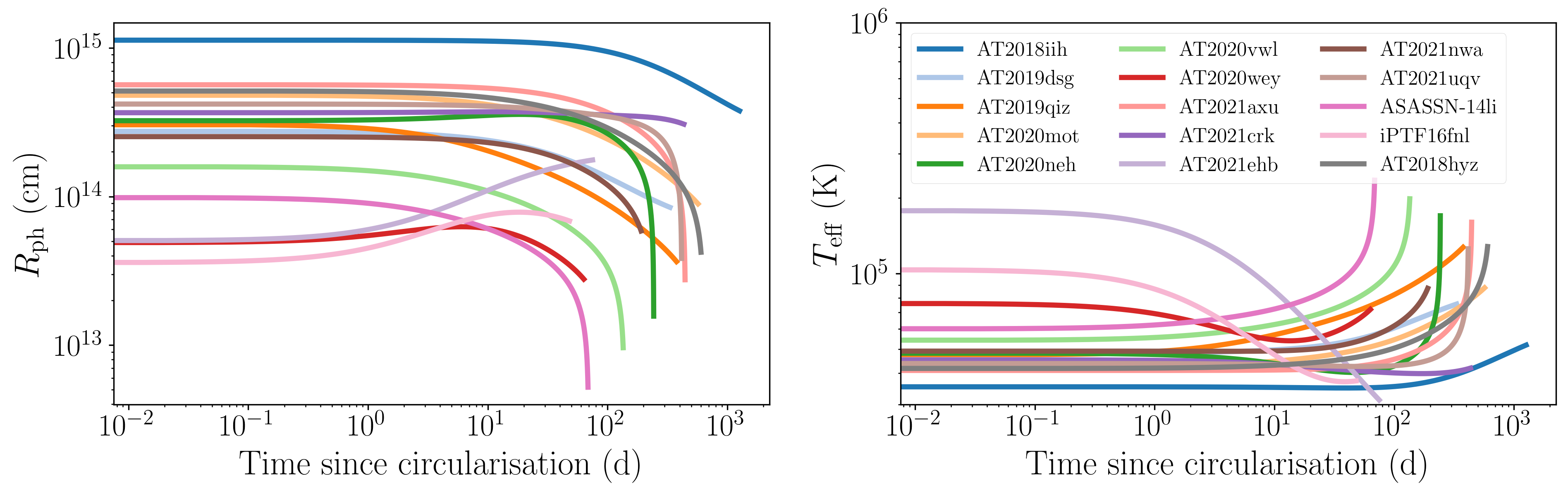} \hspace{-.4cm}
\caption{Photospheric radius and temperature evolution for time since envelope circularization for the best fit model for all TDEs in our sample.} 
\label{fig:TandRevolution}  
\end{figure*}

In Fig~\ref{fig:TandRevolution}, we show the predicted photospheric radius and temperature evolution of the best fit models for all TDEs in our sample.

\begin{table*}
\centering
\caption{Comparison between the black hole and stellar masses inferred in this work to those from the TDE sample previously analyzed with \program{TDEmass} \citep{Ryu+20} and \program{MOSFiT} \citep{Hammerstein+23}.}
\label{tab:comparison}
\begin{tabular}{lllllll}
\toprule
          & \multicolumn{2}{c}{This work} & \multicolumn{2}{c}{\program{TDEmass}} & \multicolumn{2}{c}{\program{MOSFiT}} \\
          & $M_{\rm BH}$ $[10^{6}~M_{\odot}]$ & $M_{\star}$ $[M_{\odot}]$ & $M_{\rm BH}$ $[10^{6}~M_{\odot}]$ & $M_{\star}$ $[M_{\odot}]$ & $M_{\rm BH}$ $[10^{6}~M_{\odot}]$& $M_{\star}$ $[M_{\odot}]$\\
\midrule
AT2018iih & \multicolumn{1}{c}{${5.13} \pm 0.06$} & \multicolumn{1}{c}{${9.49} \pm 0.25$} & \multicolumn{1}{c}{$6.8\pm0.2$} & \multicolumn{1}{c}{$75^{+14}_{-13}$} & \multicolumn{1}{c}{$2.2^{+0.3}_{-0.2}$} & \multicolumn{1}{c}{$3.98^{+1.51}_{-0.88}$} \\
AT2019dsg & \multicolumn{1}{c}{${1.18}_{-0.19}^{+0.18}$} & \multicolumn{1}{c}{${0.49}_{-0.06}^{+0.07}$}& \multicolumn{1}{c}{$1.3\pm0.1$} & \multicolumn{1}{c}{$3.8\pm1.0$} &\multicolumn{1}{c}{$7.2^{+1.7}_{-1.2}$}& \multicolumn{1}{c}{$8.71^{+3.04}_{-6.46}$}\\
AT2019qiz & \multicolumn{1}{c}{${1.73}_{-0.10}^{+0.11}$} & \multicolumn{1}{c}{${0.31} \pm 0.02 $}&\multicolumn{1}{c}{$1.5\pm0.1$}& \multicolumn{1}{c}{$0.65\pm0.01$}&\multicolumn{1}{c}{$2.04^{+21.4}_{-1.02}$}&\multicolumn{1}{c}{$3.00^{+0.57}_{-0.76}$}\\
AT2020mot & \multicolumn{1}{c}{${2.32}_{-0.18}^{+0.15}$} & \multicolumn{1}{c}{${1.13}_{-0.09}^{+0.10}$} & \multicolumn{1}{c}{$3.2^{+0.5}_{-0.2}$} & \multicolumn{1}{c}{$1.1^{+0.17}_{-0.10}$} &\multicolumn{1}{c}{$4.68^{+2.24}_{-1.66}$} & \multicolumn{1}{c}{$1.01^{+1.50}_{-0.12}$} \\
AT2020wey & \multicolumn{1}{c}{${0.08} \pm 0.02 $} & \multicolumn{1}{c}{${0.09}_{-0.02}^{+0.05}$} &\multicolumn{1}{c}{$0.43$} & \multicolumn{1}{c}{$0.48^{+0.02}_{-0.01}$} &\multicolumn{1}{c}{$22.9^{+3.1}_{-1.5}$} & \multicolumn{1}{c}{$4.34^{+1.96}_{-1.53}$}\\
\bottomrule
\end{tabular}
\end{table*}

\section{Sensitivity to the pre-cooling envelope phenomenology and other model assumptions}\label{sec:modelsensitivity}
\begin{figure*}[ht!]
\centering
\includegraphics[width=0.95\textwidth]{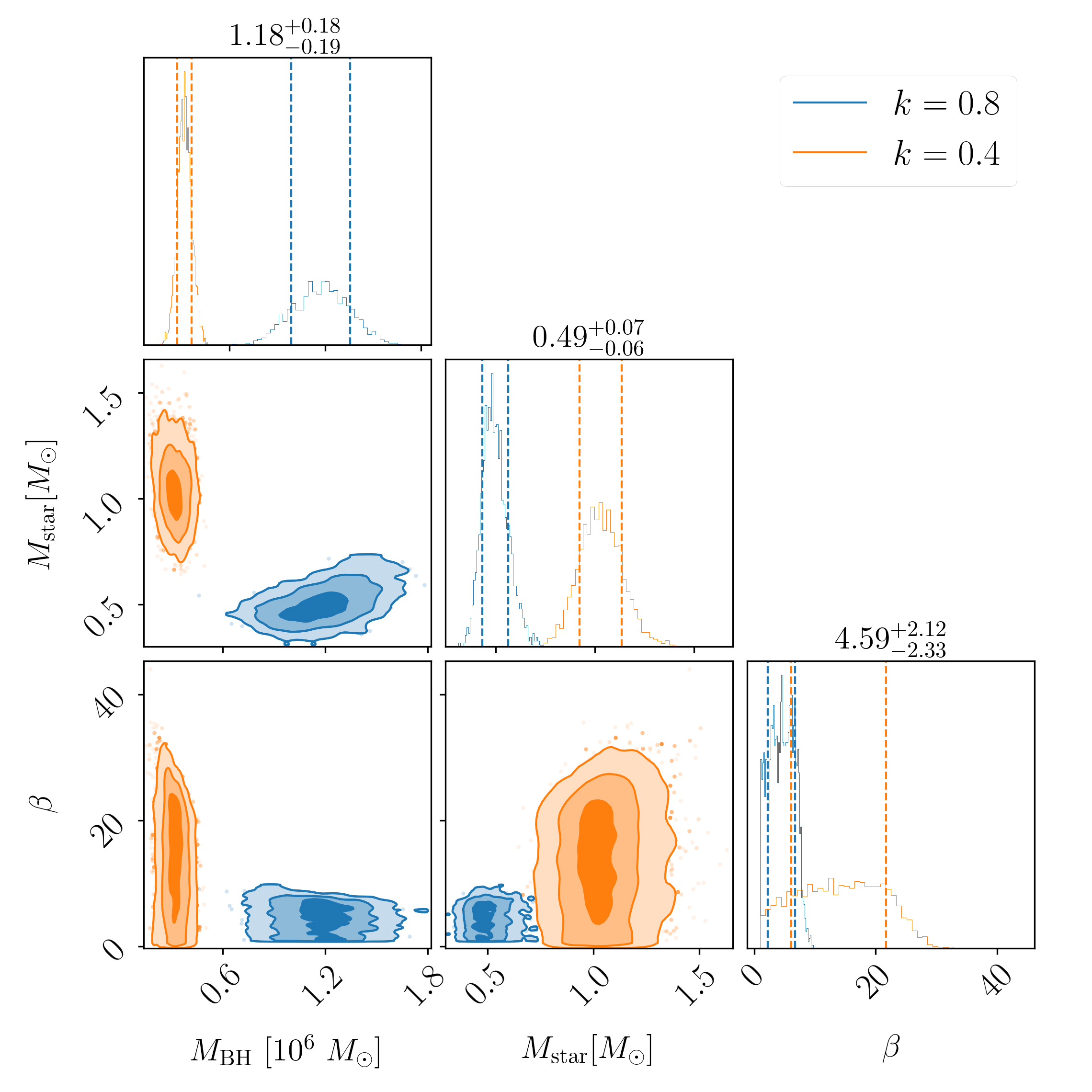} \hspace{-.4cm}
\caption{Corner plot showing the constraints for AT2019dsg for two different assumptions about the binding energy coefficient $k$ which enters the fall-back time (Eq.~\eqref{eq:tfb}), with $k = 0.8$ in blue and $k = 0.4$ in orange.} 
\label{fig:kappa}  
\end{figure*}

Here, we explore the sensitivity of our results (i.e., the inferred cooling-envelope parameters) to our choice of treatment of the pre-cooling envelope phase phenomenology and our assumption of a fixed binding energy constant, $k$. Fig.~\ref{fig:kappa} shows how our results for AT2019dsg with the preferred gaussian rise pre-cooling envelope phenomenology change if we assume $k=0.4$ instead of the fiducially assumed value $k = 0.8$. As expected, we generally find that the choice of $k$, affects the estimates of $\beta$, $M_{\rm BH}$ $[10^{6}~M_{\odot}]$, and $M_{\star}$ $[M_{\odot}]$, while the other cooling envelope parameters are largely unaffected (as they are relatively unconstrained from the prior). For AT2019dsg in particular, we note that the $k=0.8$ model provides a significantly better fit to the observations, with a Bayes factor of $\sim 10^{17}$ relative to the $k=0.4$ fit. This strong preference for a given value of $k$ supports the idea that $k$ can be added as a free parameter (or with its expected stellar-mass dependence included) in future treatments of the cooling envelope model. This in turn, would allow for a more robust estimate of the inferred masses and $\beta$.

\begin{figure*}[ht!]
\centering
\includegraphics[width=0.95\textwidth]{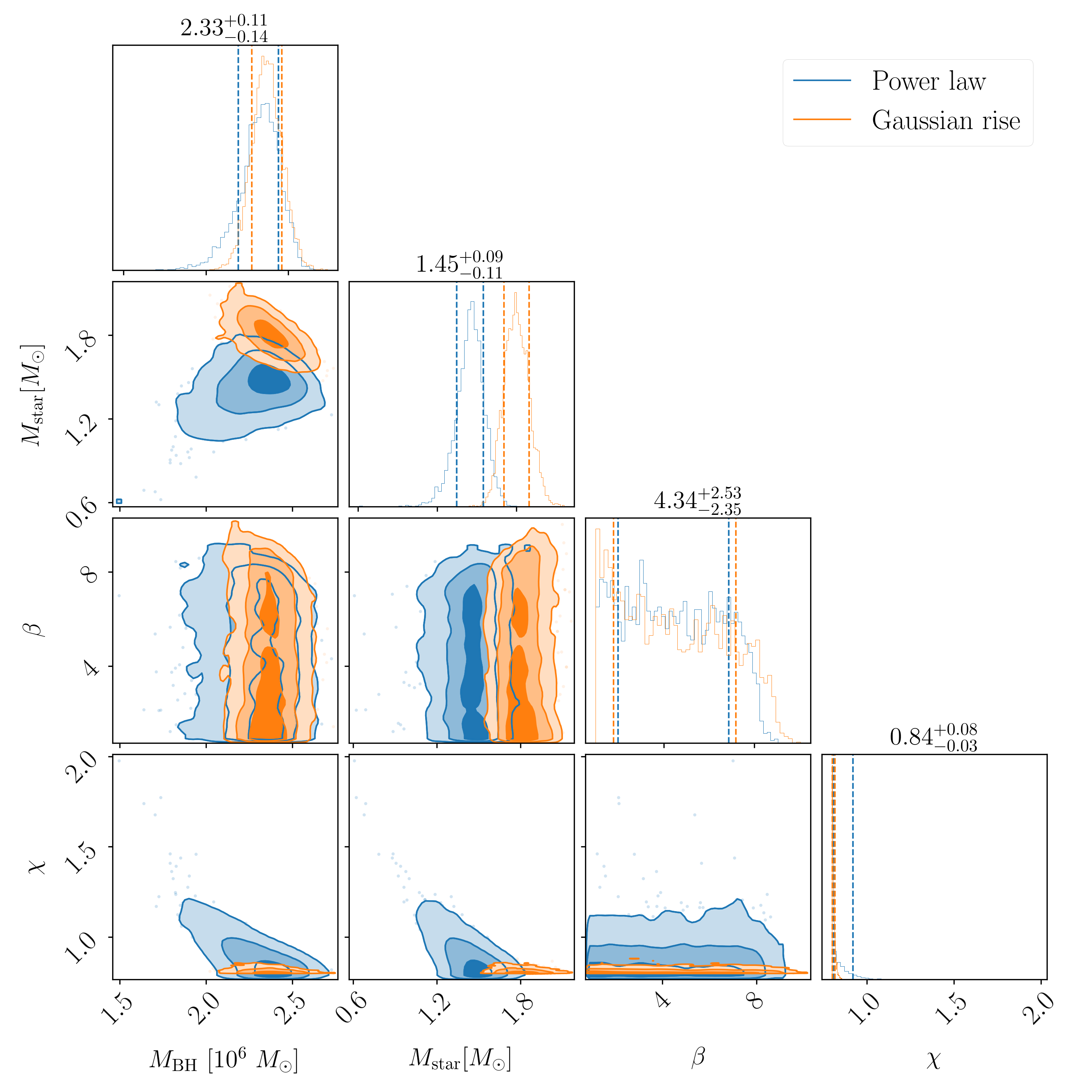} \hspace{-.4cm}
\caption{Corner plot showing the constraints for AT2018hyz for two fits with the Gaussian rise and power law pre-cooling envelope phenomenology in orange and blue, respectively.} 
\label{fig:sensitivity}  
\end{figure*}

We also have explored the sensitivity of our results to the light curve treatment prior to the cooling envelope phase. We fit all TDEs in our sample assuming both the gaussian rise and power law phenomenology. In general, we find that provided both models generate a smooth transition between the pre- and post-cooling envelope phase, the constraints so placed on cooling envelope parameters are relatively robust to this choice, with significant overlap in the posterior. Fig.~\ref{fig:sensitivity} illustrates this explicitly by comparing the constraints on AT2018hyz obtained for the two pre-cooling envelope treatments.

\bibliographystyle{aasjournal} 
\bibliography{ms}
\end{document}